\documentclass{aa}  

\usepackage{graphicx}
\usepackage{txfonts}

\begin{document}

   \title{The Gaia-ESO Survey: Detailed Abundances in the Metal-poor Globular Cluster NGC 4372}


   \author{I. San Roman\inst{1,2}\fnmsep\thanks{e-mail: isanroman@cefca.es}, C. Mu\~{n}oz\inst{1}, D. Geisler\inst{1}, S. Villanova\inst{1} , N. Kacharov\inst{3}, A. Koch\inst{3}, G. Carraro\inst{4}, G. Tautvai\v{s}iene\inst{5}, A. Vallenari\inst{6}, E.~J. Alfaro\inst{7}, T. Bensby\inst{8}, E. Flaccomio\inst{9},  P. Francois\inst{10}, A.~J. Korn\inst{11}, E. Pancino\inst{12,13}, A. Recio-Blanco\inst{14}, R. Smiljanic\inst{15}, M. Bergemann\inst{16}, M.~T. Costado \inst{7}, F. Damiani\inst{9},  U. Heiter\inst{11}, A. Hourihane\inst{16}, P. Jofr\'e\inst{16}, C. Lardo\inst{17}, P. de Laverny\inst{14},  T. Masseron\inst{16},  L. Morbidelli\inst{18},  L. Sbordone\inst{19,20}, S.~G. Sousa\inst{21,22}, C.~C. Worley\inst{16}, S. Zaggia\inst{6}}

   \institute{Departamento de Astronom\'{i}a, Casilla 160-C, Universidad de Concepci\'{o}n, Concepci\'{o}n, Chile
 \and
Centro de Estudios de F\'{i}sica del Cosmos de Arag\'{o}n (CEFCA), Plaza San Juan 1, E-44001 Teruel, Spain
\and
Landessternwarte, Zentrum f\"{u}r Astronomie der Universit\"{a}t Heidelberg, K\"{o}nigstuhl 12, D-69117 Heidelberg, Germany
\and
ESO, Alonso de Cordova 3107, 19001, Santiago de Chile, Chile
\and
Institute of Theoretical Physics and Astronomy, Vilnius University, Go\v{s}tauto 12, Vilnius 01108, Lithuania
\and
INAF - Padova Observatory, Vicolo dell'Osservatorio 5, 35122 Padova, Italy
\and
Instituto de Astrof\'{i}sica de Andaluc\'{i}a-CSIC, Apdo. 3004, 18080, Granada, Spain
\and
Lund Observatory, Department of Astronomy and Theoretical Physics, Box 43, SE-221 00 Lund, Sweden
\and
INAF - Osservatorio Astronomico di Palermo, Piazza del Parlamento 1, 90134, Palermo, Italy
\and
GEPI, Observatoire de Paris, CNRS, Universit\'e Paris Diderot, 5 Place Jules Janssen, 92190 Meudon, France
\and
Department of Physics and Astronomy, Uppsala University, Box 516, SE-751 20 Uppsala, Sweden
\and
INAF - Osservatorio Astronomico di Bologna, via Ranzani 1, 40127, Bologna, Italy
\and
ASI Science Data Center, Via del Politecnico SNC, 00133 Roma, Italy
\and
Laboratoire Lagrange (UMR7293), Universit\'e de Nice Sophia Antipolis, CNRS,Observatoire de la C\^ote d'Azur, CS 34229,F-06304 Nice cedex 4, France
\and
Department for Astrophysics, Nicolaus Copernicus Astronomical Center, ul. Rabia\'{n}ska 8, 87-100 Toru\'{n}, Poland
\and
Institute of Astronomy, University of Cambridge, Madingley Road, Cambridge CB3 0HA, United Kingdom
\and
Astrophysics Research Institute, Liverpool John Moores University, 146 Brownlow Hill, Liverpool L3 5RF, United Kingdom
\and
INAF - Osservatorio Astrofisico di Arcetri, Largo E. Fermi 5, 50125, Florence, Italy
\and
Millennium Institute of Astrophysics, Av. Vicu\~{n}a Mackenna 4860, 782-0436 Macul, Santiago, Chile
\and
Pontificia Universidad Cat\'{o}lica de Chile, Av. Vicu\~{n}a Mackenna 4860, 782-0436 Macul, Santiago, Chile
\and
Centro de Astrof\'isica, Universidade do Porto, Rua das Estrelas, 4150-762 Porto, Portugal
\and
Departamento de F\'isica e Astronomia, Faculdade de Ci\^encias, Universidade do Porto, Rua do Campo Alegre, 4169-007 Porto, Portugal
\and
Institute of Astronomy, University of Cambridge, Madingley Road, Cambridge CB3 0HA, United Kingdom
}
 
    \titlerunning {Chemical Abundances of NGC4372} 
    \authorrunning {San Roman et al.}


 
  \abstract
  {We present the abundance analysis for a sample of 7 red giant branch stars  in the metal-poor globular cluster NGC 4372 based on UVES spectra acquired as part of the Gaia-ESO Survey. This is the first extensive study of this cluster from high resolution spectroscopy. We derive abundances of O, Na, Mg, Al, Si, Ca, Sc, Ti, Fe, Cr, Ni, Y, Ba, and La. We find a metallicity of [Fe/H] = -2.19 $\pm$ 0.03 and find no evidence for a metallicity spread. This metallicity makes NGC 4372 one of the most metal-poor galactic globular clusters. We also find an $\alpha$-enhancement typical of halo globular clusters at this metallicity. Significant spreads are observed in the abundances of light elements. In particular we find a Na-O anti-correlation. Abundances of O are relatively high compared with other globular clusters. This could indicate that NGC 4372 was formed in an environment with high O for its metallicity. A Mg-Al spread is also present which spans a range of more than 0.5 dex in Al abundances. Na is correlated with Al and Mg abundances at a lower significance level. This pattern suggests that the Mg-Al burning cycle is active. This behavior can also be seen in giant stars of other massive, metal-poor clusters. A relation between light and heavy \textit{s-}process elements has been identified. 
}


   \maketitle
%

\section{Introduction}

Globular clusters (GCs) provide an exceptional laboratory for studying the star
formation history of a galaxy because they are powerful tracers of the various
components (halo, thick disk and bulge). Moreover, the ages, abundances and
kinematics of these objects represent the fossil record of the galaxy
formation process, since they are amongst the oldest objects in the Universe. Starting several decades ago, spectroscopy as well as  photometry \citep[e.g.][]{Piotto2007,Gratton2012} have shown that GCs are much more complex
than previously imagined, in particular with the almost ubiquitous discovery of multiple populations. None of the scenarios proposed to date can fully account for the abundance trends observed in Galactic Globular Clusters (GGCs). Detailed chemical studies of  large samples of GGCs are required to uncover the evolution of these objects. Providing high quality data on more GCs is clearly needed
to strengthen any conclusion. We present here results of  the first ever high-resolution spectroscopic analysis of the globular cluster NGC 4372 based on observations from the Gaia-ESO survey \citep[GES, ][]{Gilmore2012,Randich2012}. 

The GES is a public spectroscopic survey using the
high-resolution multi-object spectrograph FLAMES on the Very Large Telescope. Targeting more than 10$^{5}$ stars, GES covers all the
major components of the Milky Way, from halo to star forming regions,
providing a homogeneous overview of the distribution of kinematics and
elemental abundances. The survey also maps a significant sample of more than
80 open clusters, covering all the accessible cluster ages and masses. 

Ensuring that GES has maximal legacy impact is of key priority. Thus the
survey has identified a suitable set of objects and fields for the calibration
of the data in terms of astrophysical parameters and abundance ratios (Pancino et al. 2012). By
including objects and fields that are a) well studied in the literature; b) in
common with other large spectroscopic surveys; and/or c) with extremely well
measured properties; the results can be put in context, compared with other
surveys and when available combine different data-sets. The GES includes GGCs as calibrators, adding some relatively unstudied clusters to the ones
already present in the ESO archive. To maximize the scientific output from the
calibrator data-set, a good balance of mandatory, known and new/interesting
GGCs has been included in the target selection. Within this context, this study is focused on the chemical abundances of NGC 4372. With limited spectroscopic data, nothing is known
about any possible abundance variations of this metal-poor halo cluster.  We present here an abundance analysis of a large number of elements 
(eg. light elements, $\alpha$-elements, iron peak elements and neutron-capture
elements) and  analyze its stellar population.

The vast majority of old GGCs studied in detail to date, with the possible exception of Rup 106 \citep{Villanova2013}, show the
chemical signatures of hosting (at least) two stellar populations. 
Several other cases like Pal 12 \citep{Cohen2004}, Ter 7 \citep{Tautvaisiene2004,Sbordone2005} or Ter 8 \citep{Carretta2014} have been proposed as potential single population globular clusters but the small sample investigated does not allow for clear conclusions. We also note that all these objects except for Rup 106 are associated to the Sgr dwarf galaxy. 
There has been much improvement on the study of multiple populations within GGCs. The main evidence for this complexity comes from the presence of chemical inhomogeneities that in most cases are limited to light elements. Light elements  like Li, C, N, O, Na, Mg, Al, are
known to (anti-)correlate. The most outstanding signature is the Na-O
anti-correlation, detected so far in every GC study \citep{Carretta2009a,Carretta2009b}, again with the likely exception of Rup 106. These anomalies have been observed
also in old, massive extra-galactic GCs in Fornax \citep{Letarte2006} and in the Large
Magellanic Cloud (LMC) \citep{Mucciarelli2009}, but not in intermediate-age LMC clusters \citep{Mucciarelli2014}. A similar feature, often observed in some
GCs but not all, is the anti-correlation between Al and
Mg \citep{Gratton2001, Carretta2009b}. It is important to mention that such abundance trends are not seen in the halo field stars.

This spectroscopic evidence has been interpreted as the signature of material processed during H-burning by high temperature proton-capture reactions (like the  Ne-Na and Mg-Al cycles). Several theoretical models have been proposed in order to describe the formation and early evolution of GCs \citep{D'Ercole2008}. 
The preferred explanation involves a self-enrichment scenario, within which
two subsequent generations of stars co-exist in globular clusters and where
the second is formed from gas polluted by processed material produced by
massive stars of the first \citep{Caloi2011}. Several sources of processed
ejecta have been proposed: the slow winds of intermediate-mass AGB stars
\citep{D'Antona2002}, fast rotating massive stars \citep{Decressin2007} and
massive binaries \citep{deMink2009}. A recent attractive alternative has been proposed by \citet{Bastian2013}, that implies only a single burst of star formation. They postulate that the polluted gas concentrates in the center and is acquired by low mass stars of the first (and only) generation via disk accretion while they are in the pre-main sequence fully convective phase. This scenario has the distinct advantage that it does not require a very large percentage of the original cluster population to have been lost. In addition, variations in heavier elements have also been found in some
massive GCs such as $\omega$ Centauri \citep{Marino2011}, M54
\citep{Carretta2010}, M22 \citep{Marino2009}, NGC 1851 \citep{Carretta2011},
Terzan5 \citep{Ferraro2009}, NGC 2419 \citep{Cohen2010}, M2 \citep{Lardo2013, Yong2014} and M75 \citep{Kacharov2013}. They are generally
thought to be the vestige of more massive primitive dwarf galaxies that merged with the Galaxy. Therefore, they have important implications for the hierarchical merging scenario of galaxy formation \citep{Joo2013}. In  this paper, we use the widely accepted terminology of first and second generation of stars referring to the unpolluted and polluted populations regardless of the formation scenario.

NGC 4372 is  an old and very metal-poor globular cluster that
despite its low degree of central concentration, has received little
attention, mostly due to reddening problems. It is located close to the Galactic disk and in a dusty area of the Southern Musca
Constellation \citep[RA = 12:25:45.40, Dec=-72:39:32.4; l=300.99$^{\circ}$, b=-9.88$^{\circ}$][2010 edition]{Harris1996}. At a Galactocentric distance of R$_{gc}$ = 7.1 Kpc, NGC 4372 is listed in the Harris catalog as having a low metallicity of [Fe/H] = -2.17 and a fairly high reddening of E(B-V) =0.39, which makes it a visually
challenging object. In addition, NGC 4372 has been claimed to be dynamically associated with NGC 2808 \citep{Casetti-Dinescu2007}.

Early photometric studies of this cluster \citep{Hartwick1973,Alcaino1974,Brocato1996} already revealed a color-magnitude
diagram characteristic of very metal-poor clusters, with a well-defined horizontal
branch that extends far to the blue, and with a large and variable
 absorption. \citet{Alcaino1991} present BVRI CCD photometry in two overlapping
fields. By comparison with theoretical isochrones, they derive an age of 15
$\pm$ 4 Gyr. In a more recent study, \citet{Rosenberg2000} present a
homogeneous  photometric catalog of 39 GGCs in the southern hemisphere. They report
for NGC 4372 a
foreground reddening of E(B-V) = 0.42 , horizontal branch (HB) level of V$_{HB}$
= 15.30 with a HB ratio (B-R)/(B+V+R) = 1.00. They explain the broadening of the
CMD sequences as a consequence of the high differential reddening probably
due to the nearby Coal-sack Nebulae. A de-reddened, narrower CMD has recently been constructed by \citet{Kacharov2014}

Spectroscopic studies of NGC 4372 are limited to medium-resolution
spectra. Using the near-infrared Ca\,{\sc ii} triplet, \citet{Geisler1995} determine a mean metal abundance of [Fe/H] = -2.10 $\pm$
0.04 from 11 giant stars. They also report a mean heliocentric radial velocity
of $v_{rad}$ = 73.2 $\pm$ 1.4 km/s. Using a similar technique,
\citet{Rutledge1997} estimate a value  of [Fe/H] = -2.03 $\pm$ 0.03 on \citet[ZW84]{Zinn1984} scale and a value of [Fe/H]= -1.88 $\pm$ 0.05  on \citet[CG97]{Carretta1997} scale. A more updated value of [Fe/H]=-2.19 $\pm$ 0.08 is provided by \citet{Carretta2009a}. This value is determined from the previously mentioned Ca\,{\sc ii} triplet data but based on a more accurate metallicity scale. No further chemical study has been performed. 

In this paper, we present the first extensive study of NGC 4372  from high-resolution spectroscopy. We perform an abundance analysis of a large number of elements and analyze the stellar population of the cluster. This paper is organized as follows: Section 2 describes the observations and data reduction and Section 3  describes the methodology used to obtain the atmospheric parameters and chemical abundances. Section 4 presents our results including analysis of iron-peak elements, $\alpha$-elements and any (anti-)correlation. Finally, Section 5 presents a summary of our main results. \\

\begin{figure}
\begin{center}

\includegraphics[width=0.9\columnwidth]{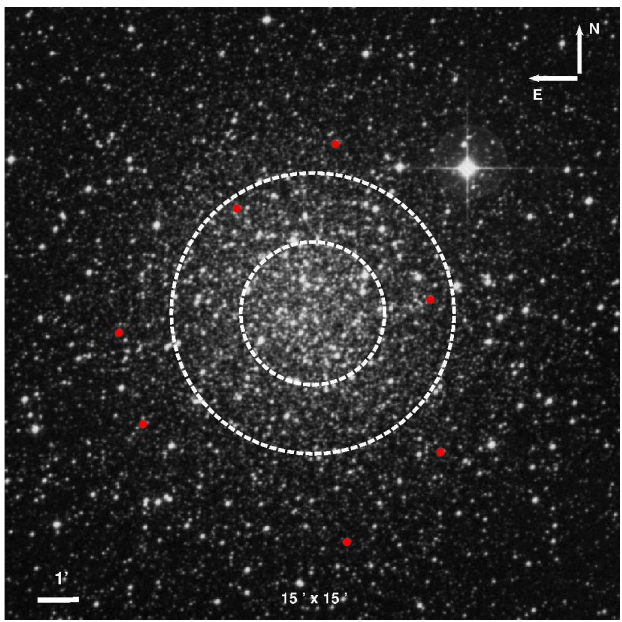}
\caption{A $15' $x $15'$ digitized sky survey image centered on NGC 4372. North is up and east is to the left. The red symbols correspond to the spatial distribution of the 7 stars analyzed. The dashed circles correspond to the core radius, r$_{c}$= 1.7' \citep[][2010 edition]{Harris1996},  and the half-light radius, r$_{h}$=3.4' \citep{Kacharov2014}.}
\label{spatial_distribution}
\end{center}
\end{figure}
\vspace*{-1cm}

\begin{figure}
\begin{center}
\includegraphics[bb=84 393 542 827, width=\columnwidth]{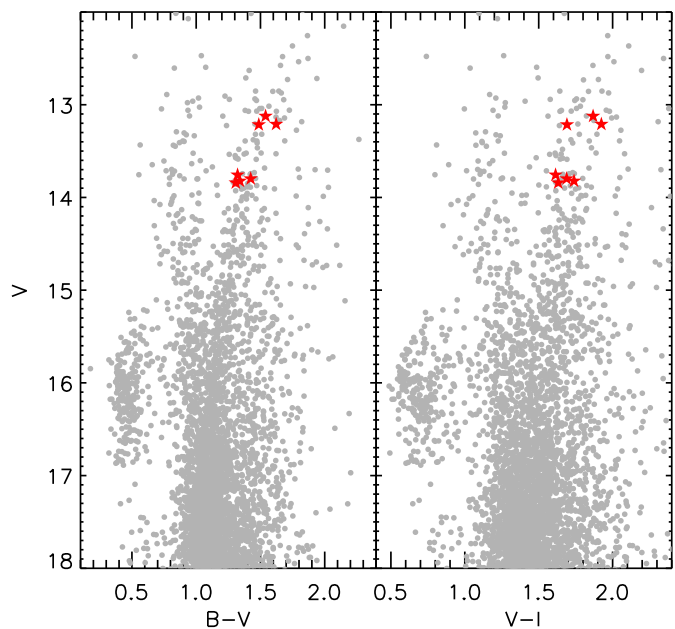}
\caption{Color-magnitude diagrams of the globular cluster NGC 4372 on an area of $6.0'$x $5.2 '$. The red symbols correspond to the analyzed RGB stars$^{1}$.} 
\label{cmd}
\end{center}
\end{figure}

\footnote{The photometric catalog has been generated by WG5 based on UBVI archival images and collected at the Wide-Field Imager (WFI) at the 2.2m ESO-MPI telescope (Programes: 164.O-0561 (PI: Krautter), 68.D-0265 (PI: Ortolani) and 69.D-0582 (PI: Ortolani)). The images were pre-reduced using IRAF package MSCRED \citep{Valdes1998}, while the stellar photometry was derived by using the DAOPHOT II and ALLSTAR programs \citep{Stetson1987,Stetson1992}. Stetson standard fields \citep{Stetson2000} have been used to photometrically calibrate the data, while astrometric calibration has been performed with the catalogs UCAC3 \citep{Zacharias2010} and GSC2 \citep{Mclean2000}. }


\section{Observations and Data Reduction}

\begin{figure}
\begin{center}
\includegraphics[bb=56 364 478 769, width=\columnwidth]{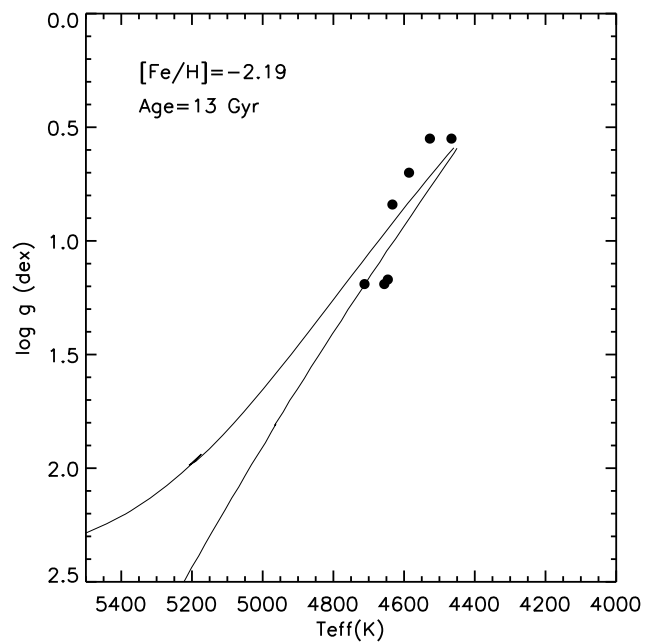}
\caption{Parameters of the stars in the $T_{eff}$- log g plane. The solid line corresponds to the isochrone computed with the web-tool of the PARSEC group \citep{Bressan2012}.}
\label{iso}
\end{center}
\end{figure}

The present work is based on the data collected by the GES from the beginning of the Survey up to the end of June 2013. These observations are referred to as \textit{gesiDR2iDR3} and are available inside the GES collaboration. As explained in \citet{Gilmore2012}, the GES consortium is structured in several working groups, WGs, having specific duties from target selection and data reduction to delivery of science data. The data reduction has been performed using a workflow specifically developed for and by GES that runs the ESO public pipeline \citep{Modigliani2004, Modigliani2012}. The pipeline has been optimized to reduce GES data and performs automatically sky subtraction, barycentric correction and normalization.  In addition the pipeline calculates radial velocities and a first guess of the rotational velocities. A quality control is performed using the output parameters from the ESO pipeline, by a visual inspection of the spectra and by the analysis of the signal-to-noise ratio of the spectra.
A detailed description of the data reduction methods can be found in  \citet{Sacco2014}. 

The present study makes use only of the UVES observations. While a small number of fibers can be dedicated to UVES, their spectra have a higher resolution than GIRAFFE and, just as important, have ten times wider spectral coverage. For this reason, UVES data are crucial for studies of precise chemical abundances of a large number of elements. As explained in Section \ref{Sec_3}, this paper makes use only of Concepcion Node atmospheric parameters and chemical abundances rather than the recommended GES values. Unfortunately, this Node does not analyze GIRAFFE spectra.  Although ideally the used of GES GIRAFFE spectra for NGC 4372 would strengthen the conclusion of this paper, the different techniques and methods employed, for this specific science case, would introduce an undesirable heterogeneity.

In this paper, we present a detailed chemical abundance analysis of seven red
giant branch (RGB) stars observed with UVES.
The stars are observed in the set-up with central wavelength 580 nm. The spectra are taken in two arms, resulting
in a wavelength coverage of 470-684 nm with a gap of $\sim$ 5 nm in the center. The
FLAMES-UVES fibers have an aperture on the sky of 1'', resulting in a
resolving power of R=47000.  The observations were taken between March 11 and
March 16 of 2012 with a mean signal-to-noise ratio in the spectrum of  $\sim$ 50. Targets for UVES were high probability cluster members and in
particular evolved stars. Figure \ref{spatial_distribution} shows the spatial distribution of the
7 RGB stars covering a wide area around NGC 4372 while Figure \ref{cmd}
presents color-magnitude diagrams of the cluster with the analyzed stars
indicated. Target selection for globular clusters has been made by WG5 (Calibrators and Standards; coordinator: E. Pancino). More details can be found in \citet{Pancino2015}.

The radial velocity information, available for all the targets observed, is determined
within WG8 by cross-correlation against real and synthetic templates. The
typical error on the radial velocities of UVES targets is about 0.4 km s$^{-1}$. The mean radial velocity value for the seven targets
is $<v_{rad}>$ = 72.6 $\pm$ 1.3 km/s while the dispersion is 3.6 km/s. The mean radial velocity is in excellent agreement with the value found by \citet{Geisler1995} of $<v_{rad}>$=73.2 $\pm$ 1.4 km/s where the error is given by the standard error of the mean.  More recently, \citet{Kacharov2014} present an extensive kinematic study of this cluster using FLAMES/GIRAFFE observations of 131 stars. They find a mean radial velocity of v$_{r}$=75.9 $\pm$ 0.3 km/s and a central velocity dispersion of  $\sigma_{0}$ = 4.5 $\pm$ 0.3 km/s. As part of the Gaia-ESO survey the kinematics of seven GGCs is present in \citet{Lardo2014}. They obtained from FLAMES/GIRAFFE spectra of more than 100 stars a mean radial velocity of v$_{r}$=75.2 $\pm$ 0.4 km/s with a velocity dispersion of 3.9 km/s, confirming the membership of our objects. The stellar parameters of stars analyzed in our study are summarized in Table \ref{Table1}.  Photometric magnitudes of the observed stars are compiled by GES and presented in Table \ref{Table_photo}.

\begin{table*}
\caption{Basic Parameters of the Observed Stars.}
\label{Table1}
\centering
\small
\begin{tabular}{cccccccc}

\hline 
\hline 
  &  Cname$^{a}$ &  v$_{rad}$   & SNR$^{b}$  & $Teff$ &  log(g)   &   $\xi$ &  $[Fe/H]$ \\
 
  &   & (km/s)   &    & (K) &  (dex)   &   (km/s) &  (dex) \\
\hline

\small         
      1  &    12250638-7243067  &    68.6   $\pm$     0.5  &   51.17  &        4586 $\pm$          86     &    0.70   $\pm$    0.20     &    1.56   $\pm$    0.10       &   -2.23  $\pm $    0.21      \\
      2  &    12250660-7239224  &    67.9   $\pm$     0.3  &   34.63  &        4646 $\pm$          56     &    1.17   $\pm$    0.15     &    1.68   $\pm$    0.03       &   -2.22  $\pm $    0.15      \\
      3  &    12253419-7235252  &    72.7   $\pm$     0.7  &   55.73  &        4527 $\pm$          33     &    0.55   $\pm$    0.07     &    1.50   $\pm$    0.17       &   -2.15  $\pm $    0.07      \\
      4  &    12253882-7245095  &    76.0   $\pm$     0.6  &   45.45  &        4656 $\pm$         114     &    1.19   $\pm$    0.47     &    1.56   $\pm$    0.20       &   -2.21  $\pm $    0.20      \\
      5  &    12260765-7236514  &    75.2   $\pm$     0.5  &   71.14  &        4466 $\pm$          38     &    0.55   $\pm$    0.09     &    1.65   $\pm$    0.05       &   -2.20  $\pm $    0.17      \\
      6  &    12264293-7241576  &    71.0   $\pm$     0.7  &   44.35  &        4712 $\pm$          87     &    1.19   $\pm$    0.21     &    1.55   $\pm$    0.13       &   -2.15  $\pm $    0.12      \\
      7  &    12264875-7239413  &    76.9   $\pm$     0.4  &   43.89  &        4633 $\pm$          68     &    0.84   $\pm$    0.34     &    1.54   $\pm$    0.05       &   -2.44  $\pm $    0.16      \\

\hline
\hline
\multicolumn{8}{p{0.75\linewidth}}{$^{a}$ GES object name. It is formed from the coordinates of the object splicing the RA in hours, minutes and seconds (to 2 decimal places) and the Dec in degrees, minutes and seconds (to 1 decimal place) together, including a sign for the declination. }\\
\multicolumn{8}{p{0.75\linewidth}}{$^{b}$ Mean signal-to-noise ratio in the spectrum.}\\
\end{tabular}
\end{table*}


\begin{table*}
\caption{Photometric Properties of the Observed Stars.}
\label{Table_photo}
\centering
\small
\begin{tabular}{ccccccccccc}

\hline 
\hline 
  &  Cname$^{a}$ & $B^{b}$ & $V^{b}$         &  $I^{b}$ & 
     $J^{c}$ &  $H^{c}$   &   $K^{c}$ &  $E(B-V)^{d}$ & $\mu_{RA}^{e}$ & $\mu_{Dec}^{e}$ \\
 
 &   & (mag)  &  (mag) & (mag)        & (mag)
 & (mag) &  (mag)   &   (mag) &  (mas/yr) & (mas/yr) \\
\hline

\small

      1  &    12250638-7243067  &   14.70  &   13.22  &   11.52  &   10.30  &    9.61  &    9.44  &    0.57  &   -6.10  &    4.80    \\
      2  &    12250660-7239224  &   15.22  &   13.80  &   12.11  &   10.86  &   10.15  &    9.98  &    0.57  &  -12.40  &    6.70    \\
      3  &    12253419-7235252  &   14.83  &   13.21  &   11.29  &    9.92  &    9.18  &    8.97  &    0.59  &   -7.10  &    8.60    \\
      4  &    12253882-7245095  &   15.15  &   13.84  &   12.21  &   11.10  &   10.43  &   10.28  &    0.54  &   -5.50  &    9.90    \\
      5  &    12260765-7236514  &   14.66  &   13.12  &   11.25  &    9.92  &    9.19  &    8.97  &    0.58  &  -13.80  &    9.00    \\
      6  &    12264293-7241576  &   15.08  &   13.76  &   12.14  &   10.98  &   10.30  &   10.16  &    0.54  &   -1.10  &    0.60    \\
      7  &    12264875-7239413  &   15.16  &   13.82  &   12.08  &   10.87  &   10.17  &    9.98  &    0.55  &   -0.10  &   13.40    \\

\hline
\hline
\multicolumn{11}{p{0.82\linewidth}}{$^{a}$ GES object name. It is formed from the coordinates of the object splicing the RA in hours, minutes and seconds (to 2 decimal places) and the Dec in degrees, minutes and seconds (to 1 decimal place) together, including a sign for the declination. }\\
\multicolumn{11}{p{0.82\linewidth}}{$^{b}$ Magnitudes from APASS (AAVSO photometric All-sky Survey).}\\
\multicolumn{11}{p{0.82\linewidth}}{$^{c}$ Magnitudes from 2MASS \citep{Skrutskie2006}.}\\
\multicolumn{11}{p{0.82\linewidth}}{$^{d}$ Reddening values from the Galactic dust extinction \citep{Schlegel1998}.}\\
\multicolumn{11}{p{0.82\linewidth}}{$^{e}$ Proper motion in RA and Dec from UCAC \citep{Zacharias2010} .}\\

\end{tabular}
\end{table*}

\begin{table*}
\caption{Abundance Ratios.}
\label{Table2}
\centering
\begin{tabular}{lcrrrrrrrrc}

\hline 
\hline 
   El.  &  Method$^{a}$ &  Nlines$^{b}$ & 1 &  2   &  3  & 4         &
   5 & 6 & 7 & Mean$^{c}$  \\
\hline

  $[O/Fe]  $   &   	SYNTH 	&  1    &		0.22       &    0.36   &    0.53       &    0.46   &    0.67       &    0.39   &    0.80       &    0.44   $\pm$    0.07         \\
  $[Na/Fe]_{NLTE}   $   & 	SYNTH	  &   2  &  	0.39       &    0.91   &    0.53       &    0.69   &    0.07       &    0.67   &    0.53       &    0.54   $\pm$    0.13            \\
  $[Mg/Fe]  $   &             EW     &  4     &             0.41       &   0.69   &    0.37       &    0.45   &    0.20       &    0.53   &    0.62       &    0.44   $\pm$    0.07          \\
  $[Al/Fe]  $   &               EW      &   2    &             0.65       &              1.16   &    0.85       &    1.12   &    ...       &    1.01   &    0.01       &    0.96   $\pm$    0.10    \\
  $[Si/Fe]  $   &               EW      &   1    &               ...       &   ...   &    0.46       &    0.52   &    0.48       &   ...   &    0.77       &   0.48   $\pm$   0.02                       \\
  $[Ca/Fe]  $   &              EW       &  18    &            0.25       &    0.29   &    0.19       &    0.24   &    0.22       &    0.28   &    0.60       &    0.24   $\pm$    0.02            \\
  $[Sc/Fe]  $   &        SYNTH       &   1   &                 -0.04       &   -0.09   &   -0.14       &   -0.14   &   -0.03       &   -0.11   &   -0.01       &   -0.09   $\pm$    0.02  \\
  $[Ti/Fe]^{d}   $   &         EW      &  14    &            0.29       &    0.40   &    0.24       &    0.30   &    0.23       &    0.38   &    0.53       &    0.31   $\pm$    0.03         \\
  $[Fe/H]  $   &                 EW      &    100  &           -2.23       &   -2.22   &   -2.15       &   -2.21   &   -2.20       &   -2.15   &   -2.44       &   -2.19   $\pm$    0.02        \\
  $[Cr/Fe]  $   &                EW     &   6    &           -0.38       &   -0.20   &   -0.33       &   -0.37   &   -0.36       &   -0.28   &   -0.15       &   -0.32   $\pm$    0.03        \\
  $[Ni/Fe]  $   &                EW       &  6    &           0.07       &   -0.03   &    0.14       &   -0.03   &   -0.06       &    0.07   &    0.06       &    0.03   $\pm$    0.03        \\
  $[Y/Fe]  $   &             SYNTH    &  1  &                 0.05       &    0.01   &    0.03       &    0.00   &    0.10       &    0.09   &    0.06       &    0.05   $\pm$    0.02     \\
  $[Ba/Fe]  $   &           SYNTH     &  1  &               -0.29       &   -0.15   &   -0.09       &   -0.16   &   -0.11       &   -0.17   &   -0.23       &   -0.16   $\pm$    0.03    \\
  $[La/Fe]  $   &            SYNTH     & 1    &                  0.09       &   -0.03   &   -0.36       &   -0.31   &   -0.06       &   -0.06   &    0.14       &   -0.12   $\pm$    0.08   \\

\hline
\hline
\multicolumn{11}{p{0.8\linewidth}}{$^{a}$ Method used in the abundance determination: SYNTH (Spectrum-synthesis method), EW (Equivalent Width method).}\\
\multicolumn{8}{l}{$^{b}$ Average number of lines employed in the abundance determination.}\\
\multicolumn{6}{l}{$^{c}$ Values obtained excluding star \#7.}\\
\multicolumn{6}{l}{$^{d}$ Values obtained from the average of Ti\,{\sc i} and Ti\,{\sc ii}. }
\end{tabular}
\end{table*}

	\begin{figure*}
\begin{center}
\includegraphics[width=0.9\textwidth]{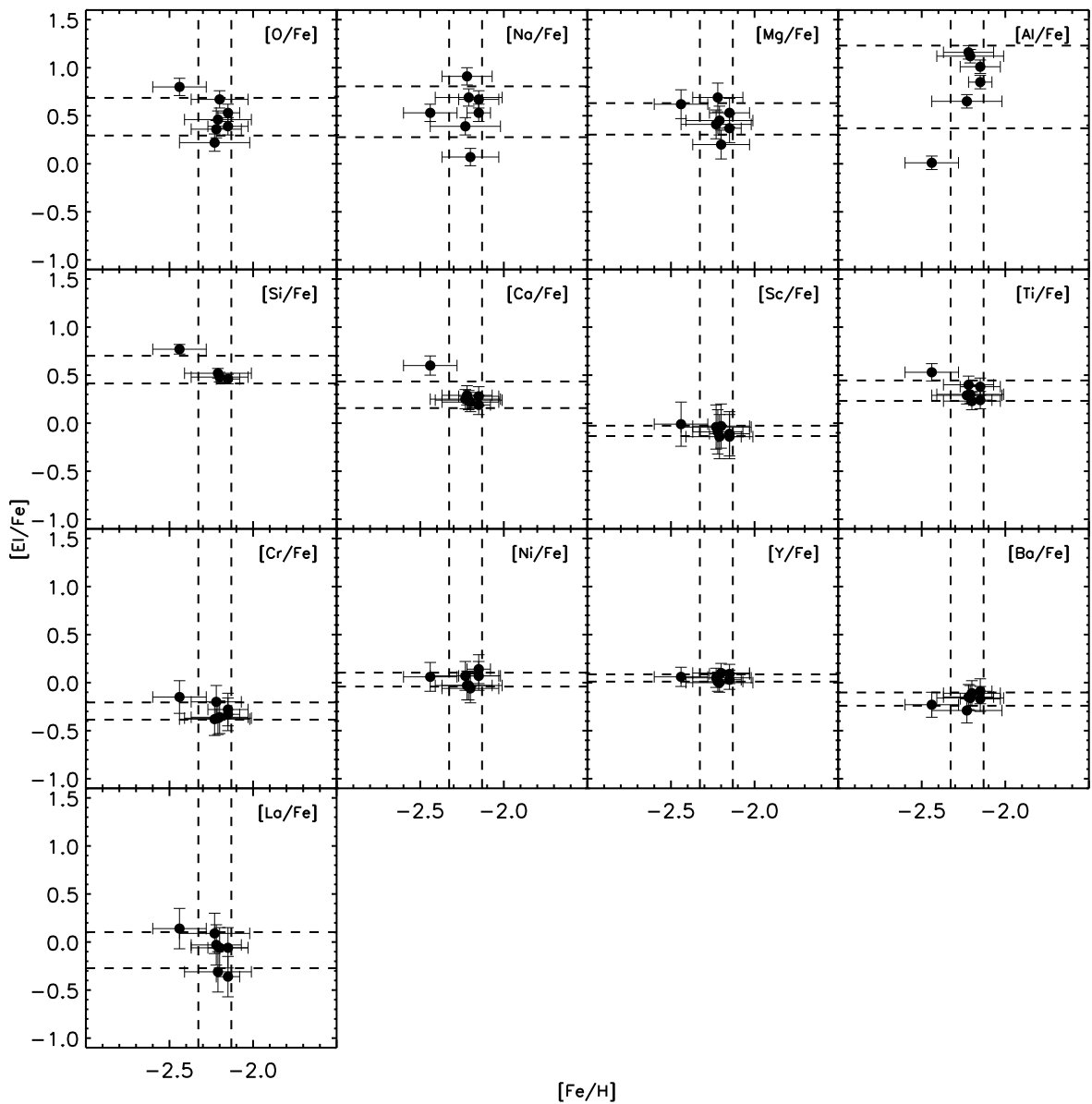}
\caption{Abundance ratios versus [Fe/H] for individual members of the
  cluster. The dashed lines indicate the $1\sigma$ area around the average value.}
\label{ratios}
\end{center}
\end{figure*}

\begin{table*}
\caption{Sensitivity of Derived Abundances to the Atmospheric Parameters and EWs.}
\label{Table3}
\centering
\begin{tabular}{ccccccc}

\hline 
\hline 
      &  $\Delta T_{eff}$=70 (k)  & $\Delta$log g=0.19(dex)   &  $\Delta \xi$=0.14 (km/s)  & $\sigma_{EW}$ &  $\sigma_{tot}$ & $\sigma_{obs}$  \\

\hline

$\Delta [O/Fe]$    &  0.04    &  0.07    &  0.03    &  0.04    &  0.09    &  0.15   \\
$\Delta [Na/Fe]$    &  0.00    & -0.07    & -0.03    &  0.04    &  0.09    &  0.29   \\
$\Delta [Mg/Fe]$    &  0.11    & -0.08    & -0.07    &  0.02    &  0.15    &  0.16   \\
$\Delta [Al/Fe]$    &  0.02    &  0.04    &  0.03    &  0.04    &  0.07    & 0.21   \\
$\Delta [Si/Fe]$    &  0.04    &  0.00    & -0.02    &  0.02    &  0.05    & 0.03   \\
$\Delta [Ca/Fe]$    &  0.08    & -0.01    & -0.06    &  0.02    &  0.10    &  0.04   \\
$\Delta [Sc/Fe]$    &  0.12    &  0.16    &  0.10    &  0.05    &  0.23    &  0.05   \\
$\Delta [Ti/Fe]$    &  0.00    &  0.07    & -0.04    &  0.05    &  0.09    &  0.07   \\
$\Delta [Cr/Fe]$    &  0.14    &  0.01    & -0.10    &  0.03    &  0.17    &  0.07   \\
$\Delta [Ni/Fe]$    &  0.13    &  0.01    & -0.08    &  0.02    &  0.15    &  0.08   \\
$\Delta [Y/Fe]$    &  0.03    &  0.07    & -0.01    &  0.06    &  0.10    &  0.04   \\
$\Delta [Ba/Fe]$    & -0.09    & -0.06    & -0.02    &  0.06    &  0.13    &  0.07   \\
$\Delta [La/Fe]$    &  0.12    &  0.14    &  0.08    &  0.07    &  0.21    &  0.17   \\

\hline
\hline
\multicolumn{5}{l}{Note.- The sensitivity determination was performed for star $\#3$.}\\
\multicolumn{5}{l}{$\sigma_{EW}$ is the error in the measurements.}\\
\multicolumn{5}{l}{$\sigma_{tot}$ is the squared root of the sum of the squares of the individual errors.}\\
\multicolumn{5}{l}{$\sigma_{obs}$ is the mean observed dispersion.}
\end{tabular}
\end{table*}

\section[]{Atmospheric Parameters and Abundance Analysis}\label{Sec_3}

The astrophysical parameters obtained from the analysis of the \textit{gesiDR2iDR3} data set will be part of the first Gaia-ESO public release of advanced data products. 
Within WG11, the spectroscopic analysis of UVES data is performed by 13 sub-groups, so-called Nodes. A multiple parallel analysis of the full data set has been implemented where different Nodes use different methodologies. Unfortunately, only a small number of Nodes reported atmospheric parameters for NGC 4372 stars. As explained in \citet{Smiljanic2014}, GES recommended parameters are computed only if at least 3 Nodes provided parameters for a given star. This decision is made based on internal policy rather than the reliability of the values. We decided to make use of Concepcion Node atmospheric parameters and chemical abundances during the following analysis. A critical evaluation of the performance of Concepcion Node, based on  a series of calibrators, is presented in \citep{Smiljanic2014}. In addition, Concepcion Node is the only one that provide, for the 7 stars observed in NGC 4372, key elements such as O or Na.

The atmospheric parameters for NGC 4372 reported by the Concepcion Node correspond to the GES iDR1 (internal data release 1) and were determined by at least 5 different Nodes in a system of multiple parallel analysis. The methodology and codes used by each Node are described in detail in Appendix A of \citet{Smiljanic2014} and they range from the classical method of equivalent width (EW) to the use of libraries of observed and/or synthetic spectra. This strategy has two main advantages: 1) Ensure that all sources of errors are well-understood and quantified, including method-dependent effects, 2) Ensure that all types of objects can be properly analyzed. To guarantee the homogeneity of the final results, a number of constraints have been imposed. These constraints include: the use of a common line list \citep{Heiter2014}, the use of one single set of model-atmospheres (the MARCS models, \citet{Gustafsson2008}), the use of a single synthetic spectrum library \citep{deLaverny2012, Recioblanco2014}, a common solar zero point \citep{Grevesse2007} and the analysis of common calibration targets. In order to understand the precision and accuracy of the atmospheric parameters, GES makes use of  the Gaia benchmark stars and a set of calibration clusters. The accuracy is judged by the ability of a given Node to recover the reference atmospheric parameters of the benchmark stars from their analysis. The precision is judged by the ability of a Node to reproduce their own results from multiple analysis of the same star.  A detail explanation of the performance of each Node as well as how the atmospheric parameters were determined can be found in \citet{Smiljanic2014}. Table \ref{Table1} summarizes the derived atmospheric parameters for NGC 4372 stars. The error reported for each parameter corresponds to the dispersion among the results from different methodologies. This method-to-method dispersion is defined as the degree to which multiple methodologies can agree on the abundance of a star. Although this dispersion is not properly the physical uncertainty of the values, it is a good indicator of the precision and is adopted as the typical uncertainty. To further investigate the reliability of these values Fig. \ref{iso} shows the derived atmospheric parameters of the stars compared with those derived from isochrones.

The chemical abundances for Mg, Al, Si, Ca, Ti, Cr and Ni were obtained using EWs of the spectral lines. The EWs were determined with the automatic code DAOSPEC \citep{Stetson2008}. GALA \citep{Mucciarelli2013} was used to determined the elemental abundances. To complete the analysis and include other key elements the abundances of O, Na, Sc, Y, Ba, and La were also obtained. For this set of elements whose lines are affected by blending, the spectrum-synthesis method were used. The local Thermodynamic Equilibrium (LTE) program MOOG \citep{Sneden1973} was used for this purpose. Five synthetic spectra having different abundances for each line were calculated, and the best-fitting value estimated as the one that minimizes the RMS scatter. Only lines not contaminated by telluric lines were used.
For both methods, the GES guidelines regarding the use of the GES line list \citep{Heiter2014}, the use of MARCS model atmospheres \citep{Gustafsson2008} and the use of \citet{Grevesse2007} solar-zero point were strictly followed.  In addition, the Na abundance was recomputed and a non-LTE correction was applied based on \citet{Mashonkina2000}. All the used Na abundances are NLTE corrected. We refer the reader to  \citet{Smiljanic2014} for a detailed description of the abundance analysis. The chemical abundances of the observed stars are presented in Table \ref{Table2}.

An internal error (star-to-star) analysis of the derived abundances has been performed to properly quantify the internal spread of abundance within a cluster.

Table \ref{Table3} lists the two primary sources of errors contributing to the total budget ($\sigma_{tot}$): the uncertainties in the measurements and the uncertainties associated with the atmospheric parameters. The parameter $\sigma_{EW}$ corresponds with the mean chemical abundance variation due to error on the EW measurements. Uncertainties in the measurements of EWs are computed by DAOSPEC \citep{Stetson2008}. These confidence intervals estimate the goodness of the fit but also take into account the quality of the spectrum (resolution, S/N, spectra defects, ...). These EW errors , $EW_{err}$, are used by GALA \citep{Mucciarelli2013} to obtain the uncertainty on the abundance of each element by varying the EW by 1$EW_{err}$.  For the elements whose abundance was obtained by spectrum-synthesis $\sigma_{EW}$ was obtained as the error given by the fitting procedure. 
An internal error analysis was performed by varying  $T_{eff}$, log(g) and $\xi$ and redetermined abundances of star $\#3$, selected as having representative atmospheric parameters. Parameters were varied by $\Delta T_{eff}$ = +70 K, $\Delta$log(g)= +0.19 and $\Delta \xi$=+0.14 km s$^{-1}$, which correspond with the mean error of the sample parameters.  We note that these parameter variations are smaller than the standard steps in the MARCS model grid. The MARCS model interpolator\footnote{The program and a detailed user manual are available on the MARCS web site: http://marcs.astro.uu.se.} has been used to generate intermediate models from the initial grid. The program interpolates the thermal structure (T), the electronic pressure (P$_{e}$), the gas pressure (P$_{g}$), the opacity ($\kappa$) and the micro turbulence velocity ($\xi$) as a function of T$_{eff}$, log(g) and metallicity [M/H]. The interpolator has been extensively tested on a previous grid of MARCS models considering the following range of parameters: 3800 K < Teff < 7000 K, 0.0 < log g < 5.0, -4.0 < [M/H] < 0.0 \citep{Masseron2006}. The interpolation is optimized to account for non-linearities in the grid and the new interpolated model must lie inside a complete cube of existing models in the parameter space (T$_{eff}$, log g, [M/H]). With the actual grid parameter steps, maximum errors in the interpolated quantities remain below 0.25$\%$ and a few $\%$ for P${_g}$ and P$_{e}$. Within these limits, we do not consider that the error budget might increase dramatically and that the mean method-to-method dispersion drives the sensitivity to atmospheric parameter variations. This estimation of the internal error was performed following the prescription of \citet{Marino2008}. The final total error, ($\sigma_{tot}$), has been computed as the square root of the sum of the squares of the individual errors. Table \ref{Table3} lists also the observed star-to-star dispersion ($\sigma_{obs}$). We remark that our goal is to search for evidence of star-to-star intrinsic abundance variation in each element by comparing the observed dispersion $(\sigma_{obs}$) and the internal errors ($\sigma_{tot}$). For this reason, external sources of error as systematics that do not affect relative abundances are not considered.

\begin{figure}
\begin{center}
\includegraphics[bb=56 363 752 861, width=1.0\columnwidth]{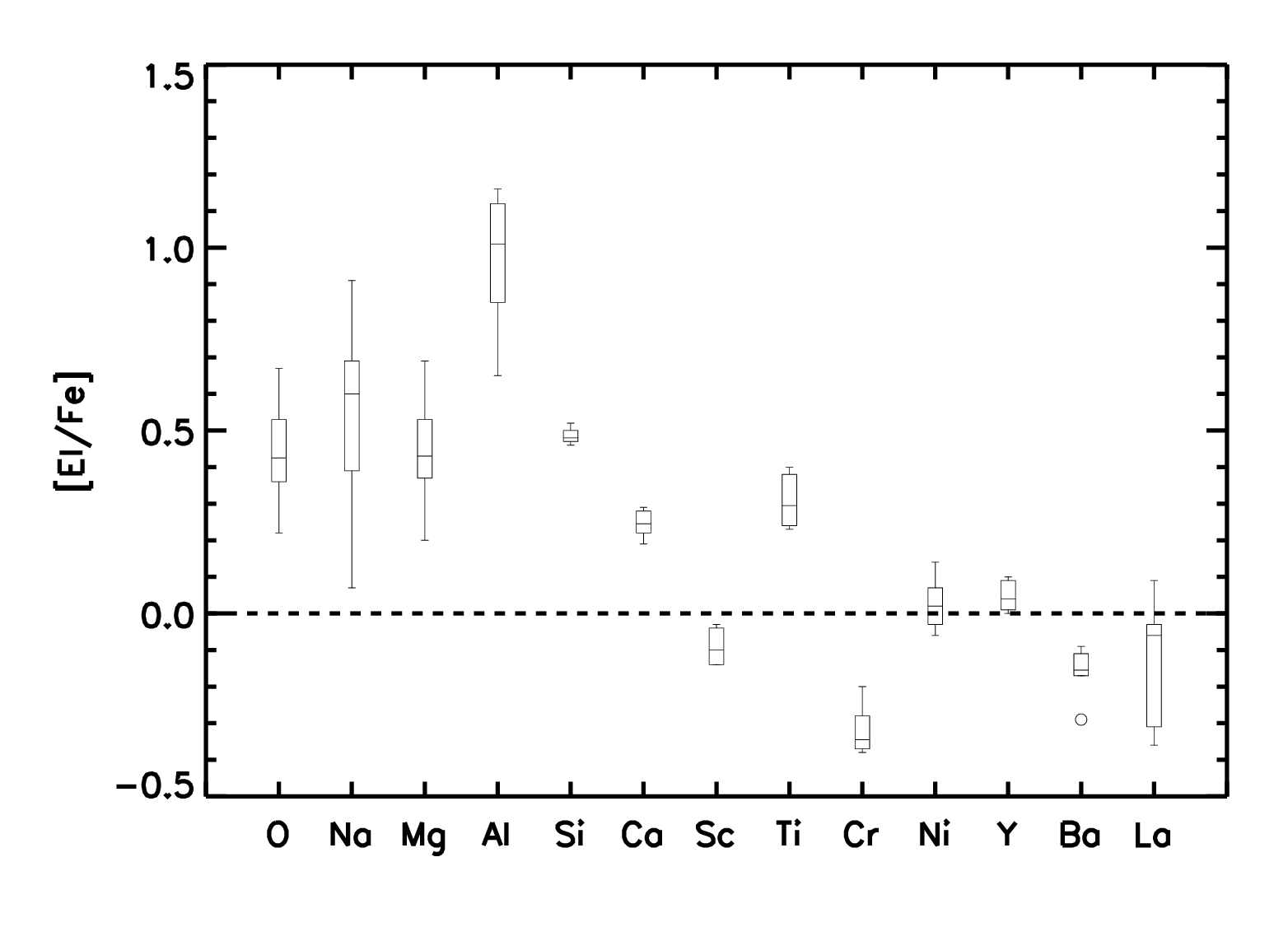}
\caption{Box plot of the NGC 4372 star element abundances. Star $\#7$ has been excluded.  For each box the central horizontal line is the median
of the data, while the lower and upper lines indicate the first and third quartile. The whiskers extend out to the maximum and minimum value of the data. Outliers are identified with small circles.} \label{boxplot}
\end{center}
\end{figure}

\section[]{Results}

	\subsection[]{Chemical homogeneity of the cluster}

	Chemically inhomogeneous populations are observed in virtually all massive, old globular clusters well-studied to date, but not in open clusters, with the possible exception of NGC 6791 \citep[][but see \cite{Bragaglia2014a} and \cite{Cunha2014}]{Geisler2012}. Studying the chemical homogeneity or inhomogeneity of a cluster is necessary to better understand the mechanism of their formation. A large body of evidence now shows that the stars in a GC do not share the same chemical composition \citep{Gratton2012}. As a general rule we can define a globular cluster as an object homogeneous in its Fe content and most other heavy elements, but the light elements Li, C, N, O, Na, Mg and Al can show substantial intracluster variations. We investigate the degree of inhomogeneity of the abundances of cluster members in NGC 4372. 
	
	Figure \ref{ratios} shows the abundance ratios versus [Fe/H] for individual member stars in the cluster. In each panel, the intersection of the dashed lines delimit a 1$\sigma$ area around the average value. Star-to-star error bar is indicated. Star $\#$7 stands out as slightly more metal poor than the main body of cluster members.  The $\alpha$-elements of this peculiar star also stand out from the cluster mean abundances, except for Mg. The radial velocity and stellar parameters for this object shown in Table \ref{Table1} are in agreement with those of the rest of the observed stars but the distinct chemical pattern of this star suggests wither that it has some kind of anomaly or it may not be a member. The possibility of star $\#$7 being in a binary system would likely alter its radial velocity as well as artificially lower its derived metallicity due to increased continuum flux. As part of the quality control, GES final products include two binary flags: a) a visual inspection of the cross-correlation function (CCF), computed before co-adding multi-epoch observations, has been performed. A star is flagged as a candidate double-lined spectroscopic binary, if the CCFs are characterized by the presence of more than one peak or a single peak with strong asymmetries; b) a star is classified as a single-lined spectroscopic binary if the median absolute deviation of multi-epoch repeated measurements of the radial velocity is larger than twice the error on the radial velocity. None of the NGC 4372 observed stars have been flagged as binary candidates.	Given its spatial location, velocity, position in the CMD and low metallicity, it seems very unlikely that this is a field star.  \citet{Kacharov2014} show that according to the prediction of the Besancon Galactic model, only a few field stars with that velocity are expected in the direction of NGC 4372 where none of them are more metal-poor than -1.8 dex. This support the view that star $\#$ 7 is a cluster member.  \citet{Lapenna2014} suggest that NLTE effects driven by over ionization mechanisms are present in the atmosphere of AGB stars and significantly affect FeI lines, while leaving FeII features unaltered. This effect could underestimate the metallicity of AGB stars. The low metallicity value of this peculiar star could be due to the fact that it is an AGB star.  It is still unclear why the $\alpha$-elements of this star are also significantly different. To be cautious and conservative, we exclude star $\#7$ from the analysis.

Fig. \ref{boxplot} presents the abundance pattern analyzed in NGC 4372. The box plot illustrates the median and the interquartile range (IQR) of the derived values. Possible outliers are also included where an outlier is defined if it deviates by more than 1.5 IQR. A large abundance ratio range is present for O, Na, Mg and Al. The star-to-star variations are smaller for the heavier elements. La presents a large abundance range but we note that the errors in the derived abundances of Sc and La are very large. For both elements, the spectrum-synthesis method with only a single line (Sc\,{\sc ii} $\lambda$5684.202 $\AA$  and La\,{\sc ii} $\lambda$5122.995 $\AA$) was used to derived the abundances. The sensitivity of the derived abundances for these two elements to the atmospheric parameters, and specifically to T$_{eff}$ and log g (see Table \ref{Table3}), is very significant and any conclusion drawn from these two elements should be treated with caution.To further investigate the degree of homogeneity of the abundances, we compare the scatter produced by internal errors, ($\sigma_{tot}$), with the observed dispersion in the chemical abundances, ($\sigma_{obs}$). These values correspond to columns 6 and 7 of Table \ref{Table3}. One can consider inhomogeneity when the intrinsic scatter is significantly higher than the expected dispersion given by the internal errors. If we exclude star $\#$7 from the error estimation, a clear intrinsic spread can be identified only in O, Na and Al.

 \begin{figure}
\begin{center}
\includegraphics[bb=61 364 490 768, width=1.0\columnwidth]{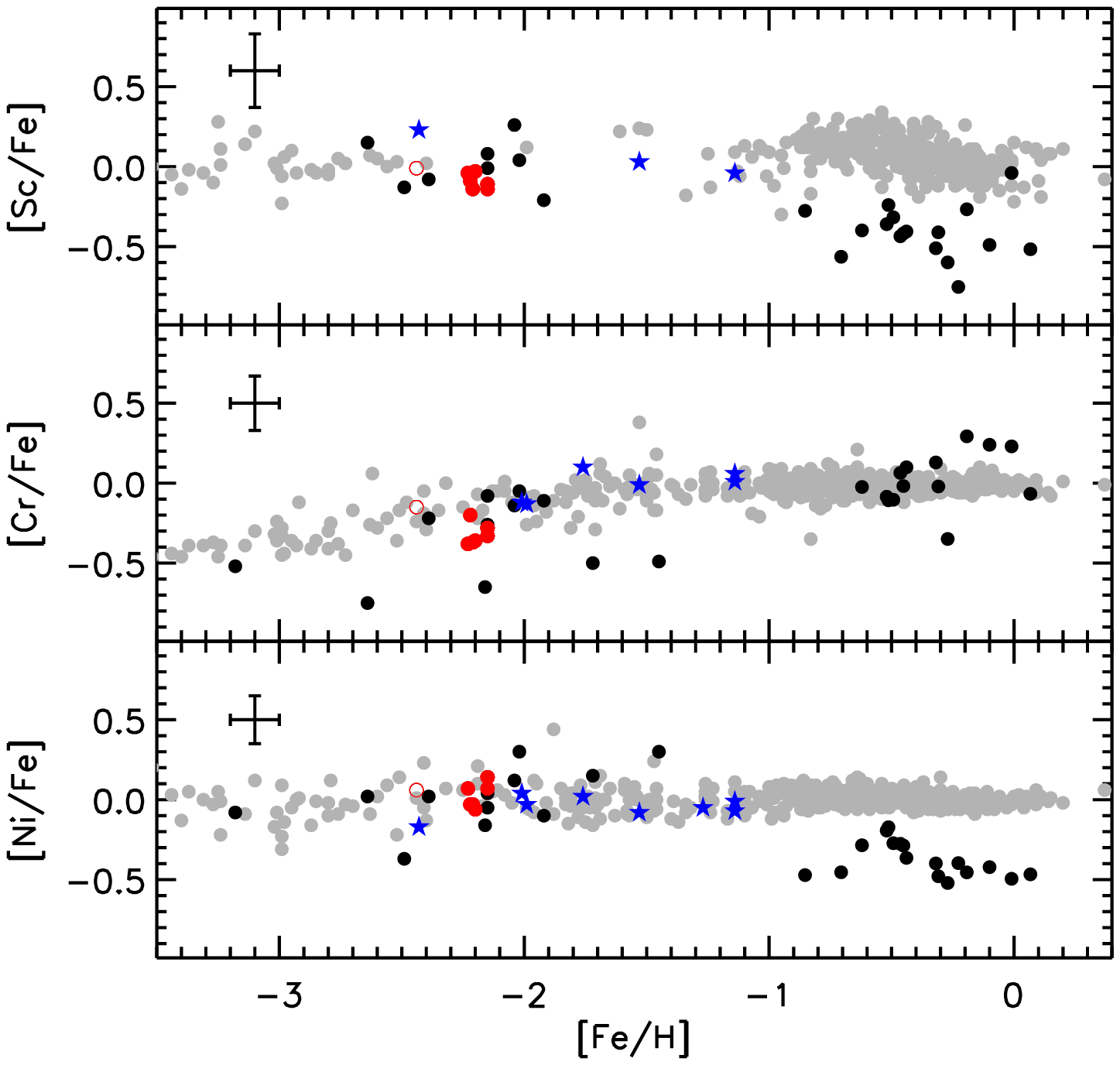}
\caption{Iron-peak element ratios ([Sc/Fe], [Cr/Fe], [Ni/Fe]) versus [Fe/H]. Red symbols correspond to values for NGC 4372 from the present study where the open red circle corresponds to the peculiar star $\#$7. Blue stars are Galactic Globular Clusters \citep{Ivans2001,Lee2005,Lee2002,Carretta2006,Carretta2009a,Carretta2010a,Villanova2011,Munoz2013, Koch2011}. Gray symbols are halo and disk stars  \citep{Fulbright2000,Reddy2003,Reddy2006,Cayrel2004} while black symbols correspond to extra-galactic objects: Draco, Sextans, Ursa minor, Carina and Sagittarius dwarf galaxy \citep{Shetrone2001,Sbordone2007, Koch2008} and the ultra-faint dwarf spheroidals Bo\"{o}tes I and Hercules \citep{Ishigaki2014,Koch2008b}. Star-to-star error bar is indicated.
}
\label{ironpeak_elem}
\end{center}
\end{figure}

\subsection[]{Iron and Iron-peak elements}
We obtain a mean metallicity for NGC 4372 of [Fe/H]= -2.23 $\pm$ 0.10. Figure. \ref{ratios} shows no evidence for an intrinsic Fe abundance spread with the exception of star $\#$7. As discussed in the previous section, this star  is anomalous in its chemical  behavior and we prefer to exclude it from cluster means. The issue of intrinsic metallicity spreads in GGCs is of great current interest. Such spreads are found generally only in the most luminous GGCs, with M$_{V}$ $\leq$ -8.5. NGC 4372 has M$_{V}$ = -7.8 so it is unlikely to host an intrinsic Fe abundance spread. However, we do note that \citet{Geisler1995} did find one of their sample of 11 stars to be significantly (about 0.5 dex) more metal-poor than the rest. This star, like star $\#$7, also had a velocity compatible with membership. Excluding star $\#$7, we found a mean [Fe/H] value of:

\begin{center}
[Fe/H]= -2.19 $\pm$ 0.03
\end{center}

The first attempt to derive a metallicity for NGC 4372 was by \citet[ZW84]{Zinn1984} obtaining a value of [Fe/H]= -2.08 $\pm$ 0.15. Most recently, \citet{Geisler1995} analyzed medium-resolution spectra of ten giant stars and obtain, through the near-infrared Ca\,{\sc ii} triplet technique, a  mean metallicity of  [Fe/H]=-2.10 $\pm$ 0.04.  Several other authors have attempted to derive the metallicity based on similar techniques but using different scales with a variety of results (eg. [Fe/H]$_{ZW84}$=-2.03; \cite{Rutledge1997}). \citet{Carretta2009c} adopt a new scale that is a fundamental shift from the older and widely used  ZW84 metallicity scale. The authors argue that this traditional scale was calibrated against only a handful of high-dispersion spectroscopic [Fe/H] values available at that time. \citet{Carretta2009c} define an accurate and updated metallicity scale using high-dispersion and high signal-to-noise spectroscopic measures of 19 GCs covering the metallicity range of the bulk of GGCs. Based on this scale they provide a value of [Fe/H]=-2.19  $\pm$ 0.08 for NGC 4372 metallicity, based on \citet{Geisler1995} Ca\,{\sc ii} triplet data. This updated result is in excellent agreement with our metallicity. The RMS scatter of our metallicity is a measure of the intrinsic spread of iron in the cluster. \citet{Carretta2009c} confirm that the scatter in Fe of most GCs is very small with an upper limit of less than 0.05 dex. Our observed scatter is consistent with that expected from errors and thus we conclude that there is no clear metallicity spread in NGC 4372, with the caveat that star $\#$7 is considered either an outlier for other reasons or a non-member.

The chemical abundances for the iron-peak elements Sc, Cr and Ni are listed in Table \ref{Table2}. The values are solar within the errors except for the Cr abundance which is underabundant. Figure \ref{ironpeak_elem} shows the elemental abundance of each star compared with a variety of galactic and extra-galactic objects. We have included values from GGCs \citep{Ivans2001,Lee2005,Lee2002,Carretta2006,Carretta2009a,Carretta2010a,Villanova2011,Munoz2013, Koch2011}; disk and halo stars \citep{Fulbright2000,Reddy2003,Reddy2006,Cayrel2004} and extra-galactic objects such as Draco, Sextans, Ursa Minor and Sagittarius dwarf galaxy and the ultra-faint dwarf spheroidals Bo\"{o}tes I and Hercules \citep{Shetrone2001,Sbordone2007, Ishigaki2014, Koch2008, Koch2008b}. 

In general, we found that NGC 4372 stars have abundances of these elements which agree with those of other GCs and halo field stars of similar metallicity. 

	\subsection[]{ $\alpha$ elements}
	
\begin{figure}
\begin{center}
\includegraphics[bb=61 365 490 855, width=1.0\columnwidth]{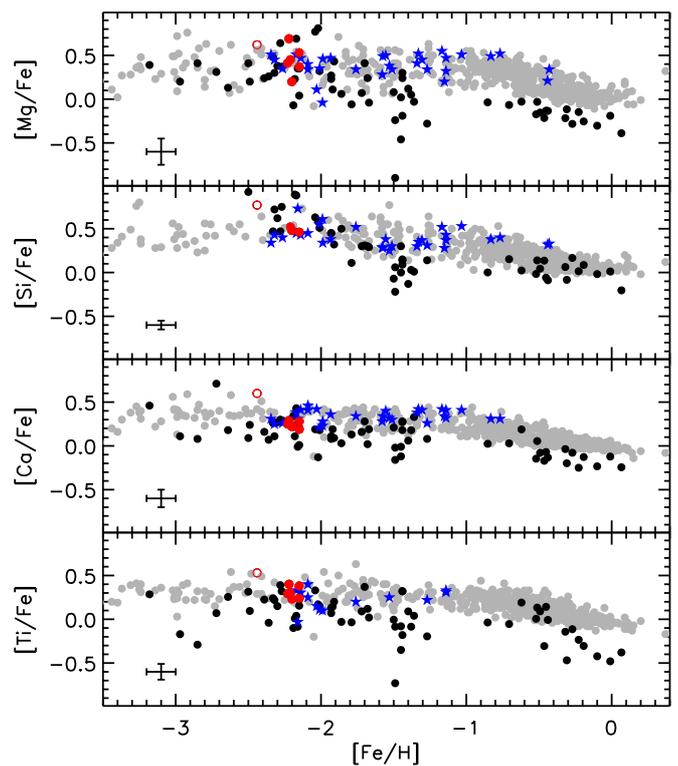}
\caption{$\alpha$-element ratios ([Mg/Fe], [Si/Fe], [Ca/Fe], [Ti/Fe]) versus [Fe/H]. Symbols are as in Fig. \ref{ironpeak_elem}.}
\label{alpha_elem}
\end{center}
\end{figure}

	All the $\alpha$ elements listed in Table \ref{Table2} (Mg, Si, Ca, Ti)\footnote{Since O shows a star-to-star variation and the Na-O anti-correlation (see Section \ref{Na_Oanti}), it will be treated separately.}  are overabundant relative to the Sun. This is a common feature among almost every GC as well as among similarly metal-poor halo field stars in the Galaxy. A glaring exception is Rup 106 \citep{Villanova2013}, which shows solar $\alpha$ element abundances. Figure \ref{alpha_elem} shows the $\alpha$-element (Mg, Si, Ca, Ti) ratios as a function of metallicity.  For comparison purpose, we have included values from GGCs, disk and halo stars and extra-galactic objects. The sources of the data are the same as those given in the previous section.  
The $\alpha$ elements in NGC 4372 seem to follow the same trend as GGCs and are fully compatible with halo field stars. Star $\#$7 stands out from the cluster behavior, suggesting once again the singularity of this star. Excluding star $\#$7, we derive for NGC 4372 a mean $\alpha$ element abundance of:

\begin{center}
$[\alpha/Fe]= +0.37 \pm 0.07$
\end{center}

\begin{figure}
\begin{center}
\includegraphics[bb=61 367 433 630, width=1.0\columnwidth]{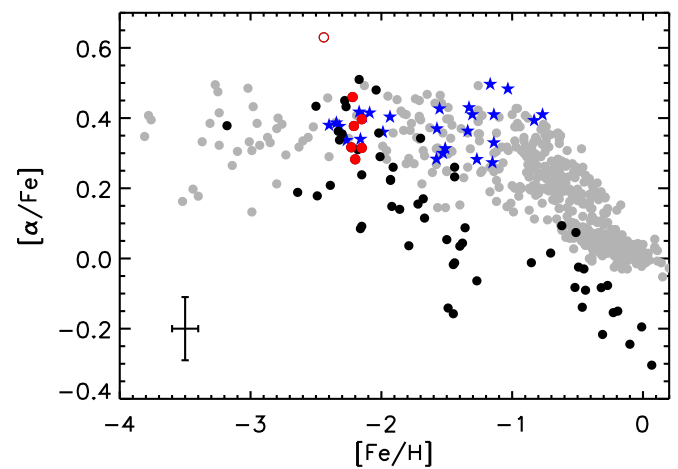}
\caption{Mean $\alpha$-element abundance versus [Fe/H]. Symbols are as in Fig. \ref{ironpeak_elem}.}
\label{alpha_mean}
\end{center}
\end{figure}

Figure \ref{alpha_mean} represents the [$\alpha$/Fe] versus [Fe/H] relation. Different symbols and colors are defined as in Figure \ref{alpha_elem}.  Similar to halo Milky Way stars, GGCs show a constant overabundance of $\alpha$ elements over a wide range of metallicities ([Fe/H] $<$ -1).  Contrarily  metal-rich dwarf spheroidal galaxies, [Fe/H] $>$ -2,  tend to have a much lower $\alpha$ content than Galactic objects at similar metallicity leading in some cases to even sub-solar ratios \citep{Geisler2007}.  NGC 4372 falls in a region where both Galactic and extraGalactic objects overlap in their $\alpha$ element content so it is not possible to draw conclusions regarding its origin from this diagram. 

\subsection[]{Na-O anticorrelation} \label{Na_Oanti}
\begin{figure}
\begin{center}
\includegraphics[bb=61 365 495 694, width=1.0\columnwidth]{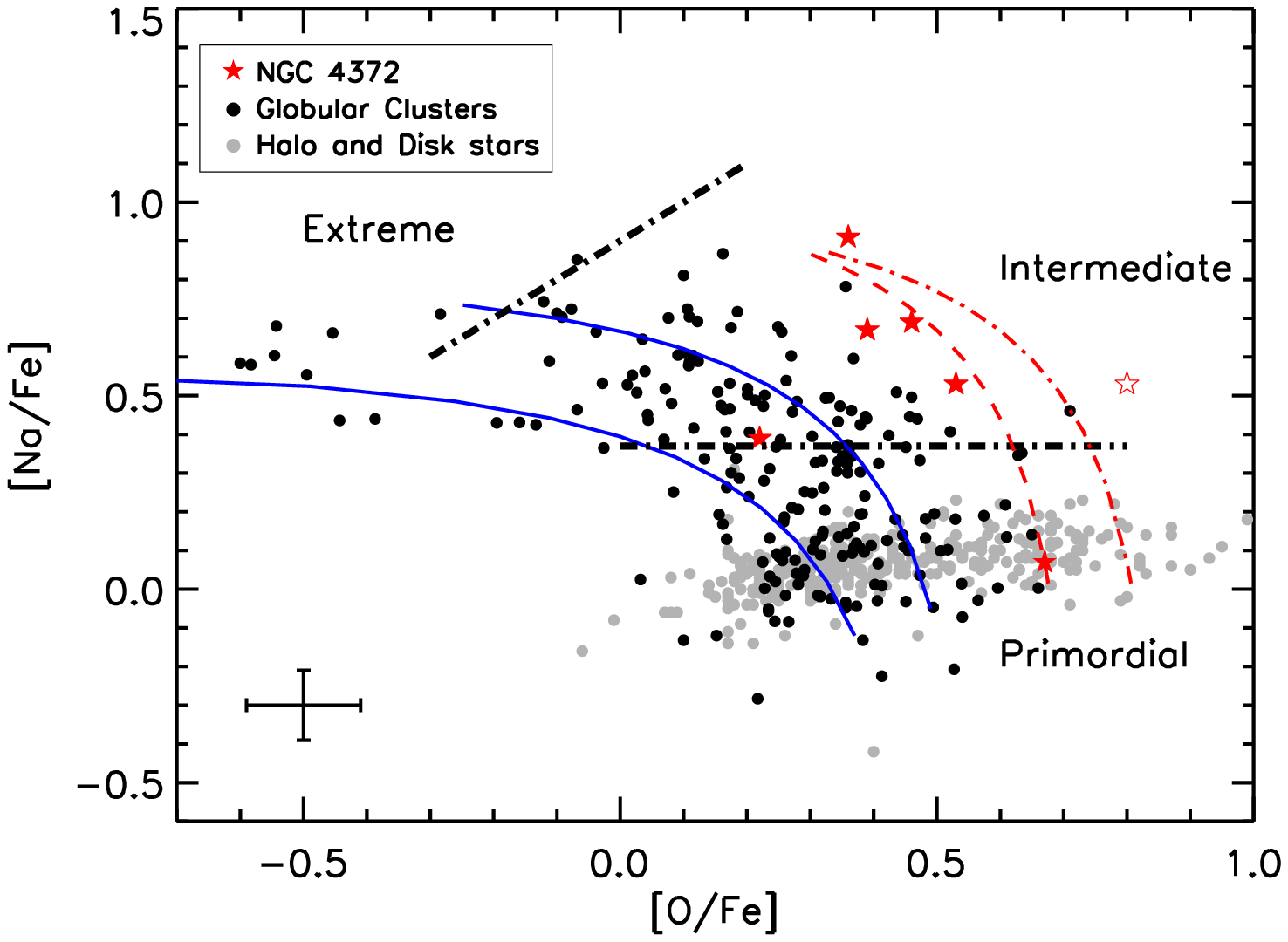}

\caption{[O/Fe] vs [Na/Fe]. Red symbols correspond with this study where the open red symbol corresponds to the peculiar star \#7. Black symbols correspond with GGC data from \citet{Carretta2009a} while gray symbols correspond with halo and disk field stars from \citet{Fulbright2000, Reddy2003, Reddy2006}. Blue solid lines correspond to the dilution model from \citet{Carretta2009b} for NGC 7078 (M15) and NGC 2808. The red dash-dotted line is the dilution model for NGC 4372 considering the full sample and the dashed line excluded star $\#7$. The black lines are the empirical separations into primordial, intermediate and extreme populations according to \citet{Carretta2009b}. Star-to-star error bar is indicated.}
\label{O_Na_correlation}
\end{center}
\end{figure}

 The Na-O anticorrelation is the classical signature of the process of proton-capture reactions in H-burning at high temperature in a previous generation of stars. All GGCs studied in detail to date show this feature with the possible exception of Ruprecht 106 \citep{Villanova2013}.  It is important to mention that such abundance trends are not seen in the halo field stars with similar metallicity. As described in the Introduction, this chemical signature has been proposed to define a GC \citep{Carretta2010b}. Figure \ref{O_Na_correlation} shows the Na-O abundances in the stars of our sample. For comparison purposes,  abundances of GGCs and halo and disk field stars have been overplotted. In addition, two distinct Na-O anticorrelations have been over-plotted as blue solid lines, which correspond to the dilution models determined by \citet{Carretta2009b}. One is O richer and is represented by the trend of the stars in NGC 7078 (M15) ([Fe/H]=-2.31)  while the other is O poorer and is represented by the trend of the stars in NGC 2808 ([Fe/H]=-1.51). Following \citet{Carretta2009b} we have obtained the dilution model for NGC 4372. The red dash-dotted line was obtained using the full sample and the dashed line excludes star $\#7$. NGC 4372 stars have a clear intrinsic dispersion and apparent anti-correlation, which corroborate the chemical inhomogeneities found in section 4.1. Figure \ref{O_Na_correlation} confirms the initial assumption that NGC 4372 is a multiple population GC. Star $\#1$ poses a value of [O/Fe] significantly smaller than the rest of the sample. This star does not follow the Na-O anti correlation describe by the rest of the stars in NGC 4372.  The O abundances were obtained using the line $\lambda$6300.30 $\AA$. This line is significantly weaker in star $\#1$ than in the rest of the sample. 
 In addition, NGC 4372 stars inhabit an area in the figure that follows the general GCs trend, although our sample lies at the high [O/Fe] end.
 
 Some interesting characteristics can be identified among our sample. One star is very Na-poor/O-rich, which corresponds to the putative primordial stellar component, while a group of Na-rich/O-poor stars would be associated  with a second generation of stars. However the low number statistics can not confirm that split. Our sample does not seem to show any star with very large O-depletion which would lie in the Extreme region, resembling the behavior of other metal-poor clusters such as NGC 7078 (M15), NGC 7099 or NGC 4590 \citep{Carretta2009b}. 

The dilution model considered by \citet{Carretta2009b} to explain the Na-O anti-correlation (Fig. \ref{O_Na_correlation}, blue and red lines), makes the basic assumptions: a) the polluting material has a well-defined composition and b) that polluting material is then diluted with a variable amount of primordial material producing the characteristic pattern of the Na-O anti-correlation. The minimum Na and maximum O abundances in each cluster represent the original Na and O composition of the cluster. NGC 4372 has abundances for O slightly high compared with the sample of \citet{Carretta2009b}. High values of O content (as an $\alpha$-element) would imply a marginal contribution by type Ia SNe to their original composition. However, the rest of the $\alpha$ abundances (Mg, Si, Ca and Ti) analyzed in NGC 4372 are full compatible with GGCs (Fig.\ref{alpha_elem} and \ref{alpha_mean}). The apparent offset of the Na-O anti-correlation observed in Fig. \ref{O_Na_correlation} could indicate that NGC 4372 was formed in an environment with high O for its metallicity.

\subsection[]{Mg-Al cycle}

\begin{figure}
\begin{center}
\includegraphics[bb=58 364 498 697, width=1.0\columnwidth]{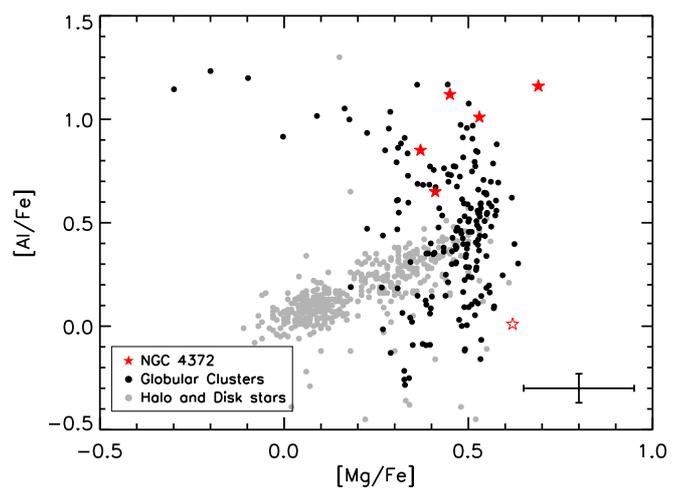}
\caption{[Mg/Fe] vs [Al/Fe]. Symbols are as in Fig. \ref{O_Na_correlation}. Star-to-star error bar is indicated.}
\label{Al_Mg_correlation}
\end{center}
\end{figure}

 \begin{figure}
\begin{center}
\includegraphics[width=1.0\columnwidth]{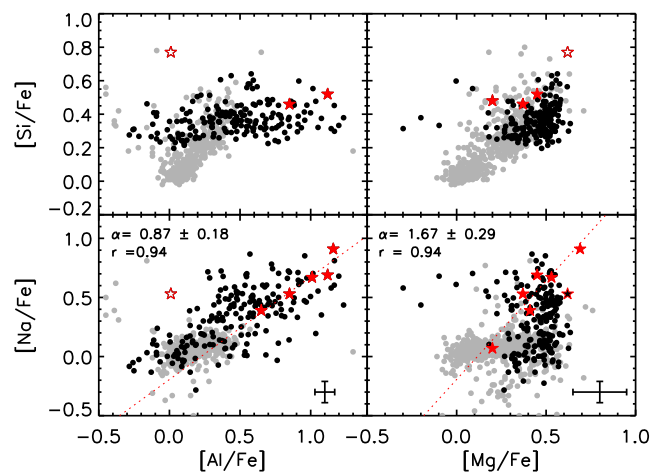}
\caption{Same as Figure \ref{Al_Mg_correlation} but for [Si/Fe] versus [Al/Fe],  [Si/Fe] versus [Mg/Fe],  [Na/Fe] versus [Al/Fe] and [Na/Fe] versus [Mg/Fe]. Symbols are as in Fig. \ref{O_Na_correlation}. Star-to-star errors bars are indicated in each panel. Dotted red lines correspond to the linear fit excluding star $\#7$. The slope of each fit is represented as $\alpha$ with the correspondent sigma error. The parameter \textit{r} corresponds to the Pearson correlation coefficient.}
\label{four_correlation}
\end{center}
\end{figure}

In addition to the CNO and Ne-Na cycle, there is evidence that the Mg-Al cycle is also active in GC polluters. Models predict that Al should show correlations with elements that are enhanced by the action of the Ne-Na (such as Na) and Mg-Al cycles and should anti-correlate with elements that are depleted in H-burning at high temperature (such as O and Mg) \citep[see discussion in][]{Gratton2012}. We analyze the abundances of some proton-capture elements to identify any (anti-)correlation in the cluster.

The pattern of abundances of the elements participating in proton-capture reactions (Na, Al, Mg, Si) observed in NGC 4372 giant stars is summarized in Figure \ref{Al_Mg_correlation} and Figure \ref{four_correlation}. The level of scatter in all these plots is significant. For comparison purposes,  abundances of GGCs and halo and disk field stars have been over-plotted. If the temperature is high enough, one should expect some degree of anti correlation between Mg and Al. However the production of Al at the expense of Mg via the MgAl-cycle does not result into such a well-defined anticorrelation as the Na-O one \citep{Langer1995}. In fact, large variations in Al are often accompanied by much smaller changes in Mg abundances \citep{Carretta2009a}. The Mg-Al behavior in NGC 4372 (Figure \ref{Al_Mg_correlation}) presents similar features to the ones found in studies of other GGCs \citep{Carretta2009a}. 

A clear Al spread is present in our sample with a star-to-star variation range in the Al abundances of $\sim$ 0.5 dex (excluding star $\#$7). However we need to consider the potential influence of NLTE effect in Al abundances.  \citet{Thygesen2014} show that there is a non negligible NLTE effect on at least some Al lines. This NLTE effect correlates with T$_{eff}$ where cooler stars would show stronger negative NLTE correction that warmer ones. Taking into account the small range of stellar parameters (and in particular T$_{eff}$) of our sample stars, the Al spread should not be affected significantly by NLTE corrections. On the other hand, the larger uncertainty on Mg suggests that Mg may be consistent with no spread. The correlation detected among stars of NGC 4372 follows the behavior displayed by GGC stars with a clear distinction from disk and halo field stars. Unlike the well-known O-Na anti correlation, the Mg-Al anti correlation is more difficult to reproduce in simulations \citep{Denissenkov1998,Ventura2001}.  \citet{Carretta2009a} show that Al-rich and Mg-depleted stars are present only in massive clusters (NGC 2808, NGC 6388, NGC 6441), metal-poor clusters (NGC 6752) or both (NGC 7078 = M15). In those clusters a clear Mg-Al anti correlation is observed even among main sequence stars \citep{Bragaglia2010}. More precise Mg values would be required to clarify the present of a Mg spread in NGC4372. The presence of a Mg-Al anti correlation in NGC4372 would indicate that the polluted generation has been enriched by material from stars where the Mg-Al burning cycle was active.

Top panels in Figure \ref{four_correlation} show the abundances of elements involved in the Ne-Na, Mg-Al cycles of proton-capture reactions in high temperature H-burning. The dotted red line in each panel of Figure \ref{four_correlation} corresponds to a linear fit of the sample (excluding star $\#7$) where the parameter \textit{$\alpha$} represents the slope of the fit with the sigma error. The Pearson correlation coefficient, \textit{r}, is also shown in each panel. \citet{Carretta2009a} found that stars with extreme Al overabundance also show Si enhancement with respect to the remaining stars in the same cluster. Again, this effect is limited to massive or metal-poor GGCs. The correlation between Si and Al abundances is a signature of production of $^{28}$Si from the Mg-Al cycle \citep{Yong2005}. The reaction producing this isotope becomes predominant in the Mg-Al cycle when the temperature is very high (exceeding T$_{6}$ = 65 K) \citep{Arnould1999}.  The chemical pattern observed in these giant stars must be imprinted by a previous generation of massive stars to be capable to reach those high temperatures.  If we exclude star $\#$7 from the analysis (open red symbol), Si does not show large star-to-star variations but unfortunately the number of stars with reliable Si abundances are very limited. A clear Na-Al correlation is evident in our data. This is a reflection of the relation between Mg-Al and Ne-Na cycles \citep[see discussion in][]{Gratton2012}. In fact, Al and Na are predicted to be simultaneously enhanced when the Ne-Na and Mg-Al cycles are both acting. A Na-Mg relation is also identified in NGC 4372 stars. This behavior does not follow the general trend of GGCs and in fact one would expect to find a Na-Mg anti correlation. This relation is driven mostly by two stars only. Large Mg errors could cause this unexpected relation. More extensive samples of stars with accurate determinations of abundances are required for a more detailed analysis.

\begin{figure}
\begin{center}
\includegraphics[bb=58 364 498 697, width=1.0\columnwidth]{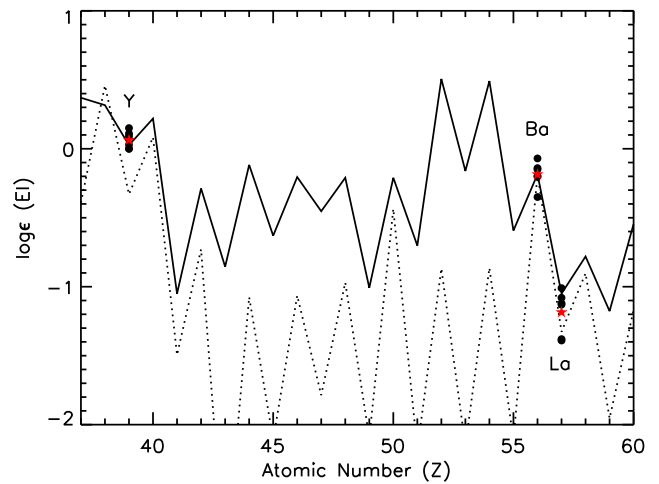}
\caption{Abundance distribution of heavy neutron-capture elements in NGC 4372. Black symbols correspond to the individual stars while red symbols represent the mean abundance for the cluster (excluding star $\#7$). Solid and dashed lines show the scaled-solar \textit{r-} and \textit{s-}process patterns respectively \citep{Simmerer2004} and have been normalized to the mean abundance of Ba. Notation: log $\epsilon$ (El)= log$_{10}$(N$_{El}$/N$_{H}$) + 12.0}
\label{sprocess}
\end{center}
\end{figure}

\begin{figure*}
\begin{center}
\includegraphics[width=0.9\textwidth]{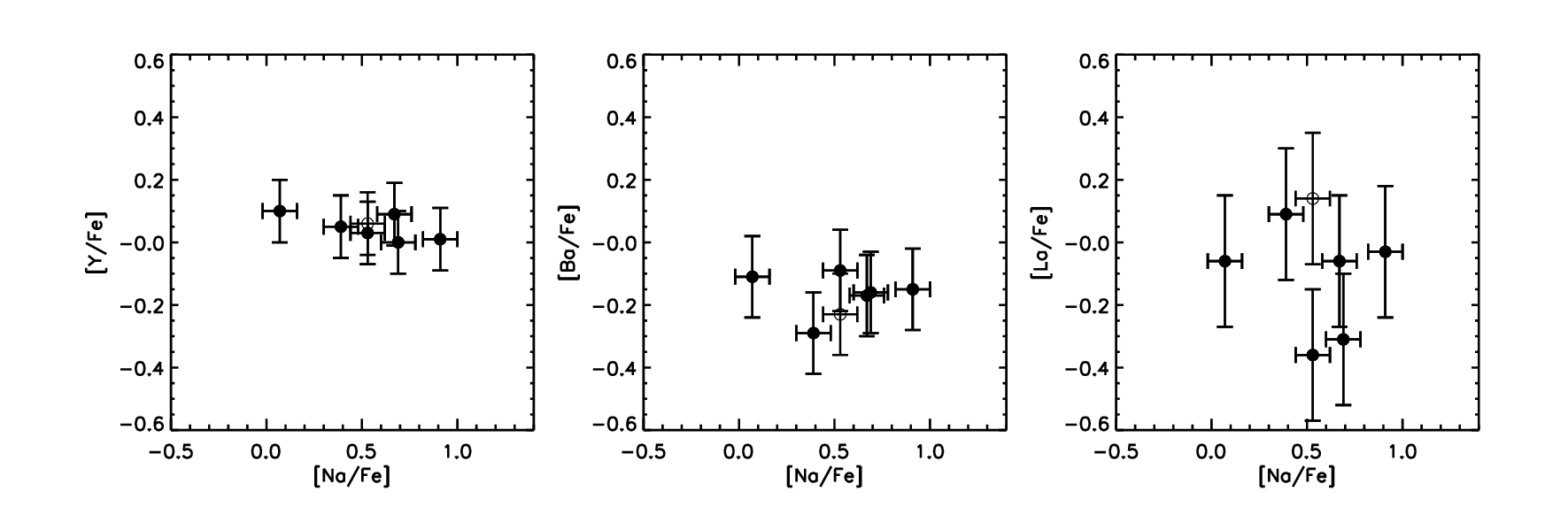}\vspace{-0.5cm}
\includegraphics[width=0.9\textwidth]{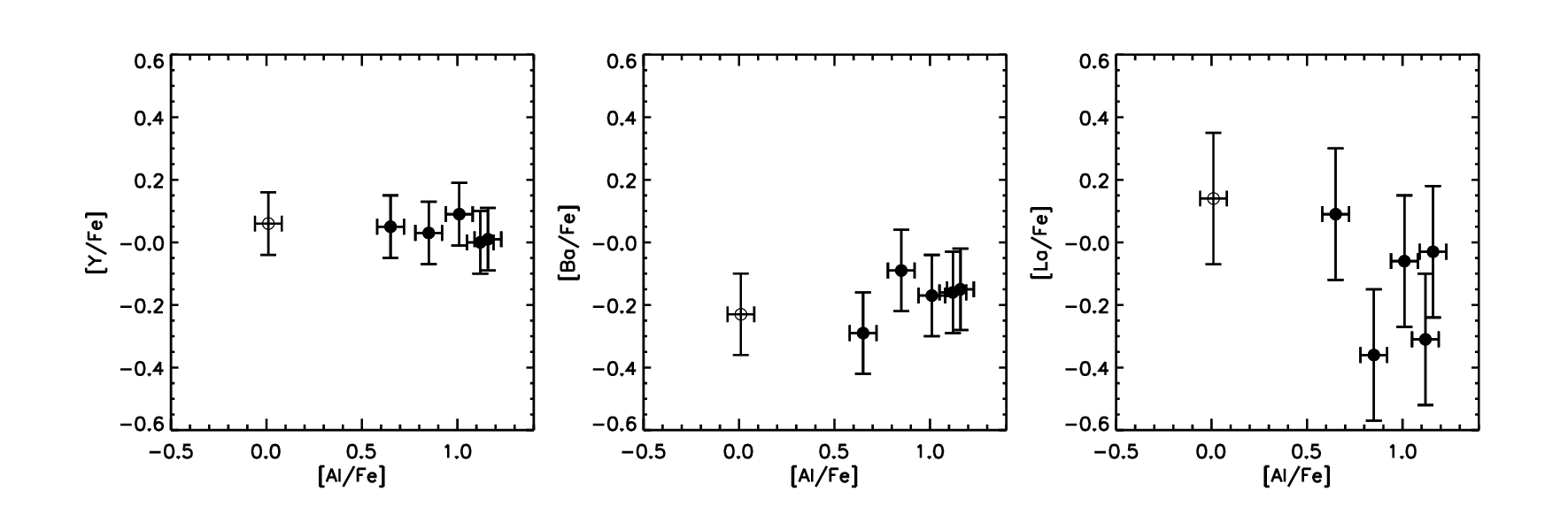}
\caption{Neutron capture elements [Y/Fe], [Ba/Fe] and [La/Fe] as a function of light elements [Na/Fe] and [Al/Fe]. Open symbol corresponds to star \#7.}
\label{heavyelements}
\end{center}
\end{figure*}

 \begin{figure*}
\begin{center}
\includegraphics[width=0.9\columnwidth]{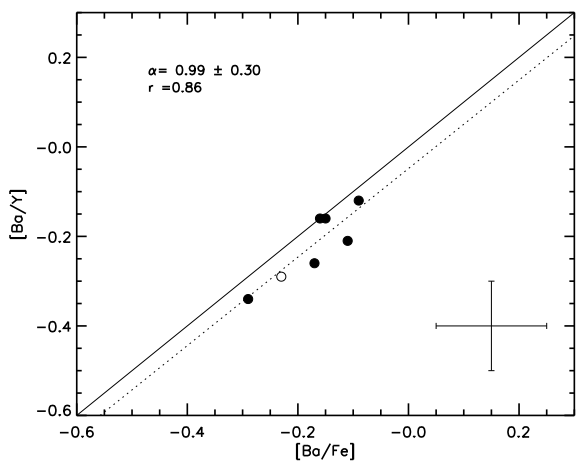}
\caption{Relation of the light and heavy \textit{s-}process elements,  [\textit{hs/ls}]=[Ba/Y], as a function of [Ba/Fe]. Open symbol corresponds to star \#7. The dashed line corresponds to the linear fit excluding star $\#7$. The solid line corresponds with the 1:1 correlation. The slope of the fit is represented as $\alpha$ with the correspondent sigma error. The Pearson correlation coefficient is also shown.}
\label{hsls}
\end{center}
\end{figure*}

\subsection{Heavy Elements}
While light-element variations are well-known in GCs, intrinsic dispersion among heavier elements is less common. Most of the heavier elements (Z $>$ 30) are produced either by slow or rapid neutron-capture reactions (the so-called \textit{s-} and \textit{r-} processes). S-process happens in a different physical condition with respect to \textit{r-}process and are thus likely to happen in different astrophysical sites.  We measured the abundances of the neutron-capture elements: Y, La and Ba. These elements are mainly produced by the \textit{s-}process at solar metallicity in which the neutron-capture time is much longer than the beta-decay lifetime.  From our data we could not determine the \textit{r-} and \textit{s-}process ratio of neutron capture, because we did not measure a reliable abundance for a typical \textit{r-}process species (e.g. Eu).  

The abundance pattern of heavy neutron-capture elements (Y, Ba, La) is shown in Fig. \ref{sprocess}. We have over plotted the pure \textit{r-} and \textit{s-}process patterns from \citet{Simmerer2004}.  The Ba abundance was used to set the zero point of the curves.  Although the analysis includes very few points for any solid conclusion, the observed abundance of Y can be reproduced by a pure \textit{r-}process pattern.

Figure \ref{heavyelements} shows the abundance ratios of the heavy elements analyzed as a function of [Na/Fe] and [Al/Fe]. The [Ba/Fe] abundance ratio is sub-solar in agreement with recent studies which found that [Ba/Fe] decreases with low metallicites and becomes sub-solar at least in fields stars with [Fe/H] $<$ -2.0 \citep[e.g.][]{Gratton1994,Fulbright2002}.  These sub-solar values will agree with recent studies that indicate that metal-poor globular clusters primarily exhibit \textit{r-}process signatures, due to the inefficiency of \textit{s-}process in low metallicity environments \citep{Gratton2004,Roederer2010, Koch2009}. At [Fe/H] < -2 dex, the environment was not yet polluted by a sufficient number of low mass AGB stars and the average \textit{s-}process abundances are low. All the elements show an apparent small dispersion however, the observed range is very comparable to that expected from the errors (see Table \ref{Table3}) so we can not confirm any clear evidence for an intrinsic spread in any of these elements.  

Only a handful of metal-poor globular clusters show a potential star-to-star dispersion in neutron-capture elements \citep{Roederer2011, Kacharov2013}.  \citet{Marino2009} found a wide range of abundances values for \textit{s-}process elements, Y, Zr and Ba, in M22. They also identified a bimodality among these elements. None of the elements show a correlation with Na, O and Al. The bimodality in \textit{s-}process elements in M22 resembles the case of NGC 1851 \citep{Yong2008} although in this case the \textit{s-}element abundance appears to correlate with the Na, Al, and O abundances.  
The metal-poor globular cluster M 15, with [Fe/H] =  - 2.31 $\pm$ 0.06 dex \citep{Carretta2009b}, has also proven to be an interesting case for study. \citet{Sneden1997, Sneden2000} found a scatter of heavy neutron-capture elements but no significant \textit{s-}process enrichment in M15. More recently, \citet{Worley2013} found a bimodal distribution of Ba, Eu and possibly La in M15. Both modes of the bimodality seem to be indicative of a pollution scenario dominated by the \textit{r-}process and only varying due to a different degrees of enrichment. A group of La-rich and La-poor stars can be identified in our sample but such a small sample does not allow us to carryout further investigation.  

 The \textit{weak} \textit{s-}process accounts for the major fraction of the light \textit{s-}process (\textit{ls-}) elements, like Y, and occurs in core He-burning massive stars  (M$ > $ 8M$_{\odot}$) \citep{Pignatari2010}. The \textit{main} \textit{s-}process takes place in thermally pulsating AGB stars (1M$_{\odot}$$ <  $M$  < $8M$_{\odot}$) producing light \textit{s-}process (\textit{ls-}) and heavy \textit{s-}process (\textit{hs-}) elements, like Ba and La \citep{Arlandini1999}. Therefore, analysis of ratios between \textit{ls-} and \textit{hs-} elements are very interesting in order to constrain these formation processes. Figure \ref{hsls} shows the [\textit{hs/ls}]=[Ba/Y] ratio as a function of the heavy \textit{s-}process element Ba.  Although the large errors should be considered, a relation is present between Y and Ba. This relation could imply a real spread of Y and Ba. Although the spread among heavy elements is comparable to the star-to-star scatter found in [Fe/H], heavy element absorption lines are generally difficult to measure. For example, Ba absorption lines are quite strong among luminous RGB stars so saturation could be a problem. They have hyperfine structure and involve the choice of atmospheric parameters. In particular, they are quite sensitive to the choice of microturbulent velocities. These effects could produce an increase in the uncertainties. 

 The \textit{s-}process elements can be considered as a signature of the processes that occur in intermediate mass AGB stars \citep{Busso2001}, whose wind could have polluted the primordial material from which the second generation of stars formed. If the spread in \textit{s-}process contents in NGC 4372 is real, this would argue against the abundances of \textit{s-}process elements being intrinsic to the cluster and suggest that AGB stars can be a possible polluter. However, none of the analyzed heavy elements show any clear relations with other lighter elements that could suggest an obvious pollution by AGB stars. Since this potential spread seems unrelated with the spreads in light elements, the variations could be attributed to the original inhomogeneities in the gas which formed the cluster. A larger sample survey of stars at various evolutionary phases and extensive abundance analysis of a range of key nucleosynthetic indicators would be essential to confirm whether this potential star-to-star scatter is real.

\section{Summary}

In this paper we present the first detailed chemical abundances of 14 elements in 7 red giant members of NGC 4372 using high resolution, high S/N spectroscopy. Chemical abundances have been computed by Concepcion Node within the GES collaboration. The classical EW method has been used when possible.  For 5 elements whose lines are affected by blending, the spectrum-synthesis method was preferred. One of the stars of our sample shows a radial velocity in agreement with the cluster bulk but a distinct chemical signature so we have excluded it from the statistical analysis. 

We found a metallicity of [Fe/H]= - 2.19 $\pm$ 0.02 with a $\sigma_{obs}$ = 0.03 dex, in good agreement with previous, low-resolution, studies. We rule out an intrinsic metallicity spread, although the low value for the one (excluded) star should be born in mind. We confirm the Na-O anti correlation although not very extended, probably due to our small sample. The abundances of O are relatively high compared with other globular clusters which could indicate that NGC 4372 was formed in an environment with high O for its metallicity. Intrinsic spreads are also seen in other light elements, in particular an apparent Mg-Al anti correlation has been detected. The Fe-peak elements generally show good agreement with other GCs and halo field stars with no dispersion. The $\alpha$ elements  show an  enhancement of [$\alpha$/Fe]= +0.37 $\pm$ 0.07  typical of other GCs indicating similar fast star formation time scales. A relation between light and heavy \textit{s-}process elements has been identified. A larger sample of stars at various evolutionary phases and extensive chemical  analysis is required for a more definitive analysis.

\begin{acknowledgements} 
We thank Michele Bellazzini for a very careful reading and useful comments and suggestions that helped to improve the quality of the paper. We also thank an anonymous referee for comments that greatly improved this paper. \\

Based on data products from observations made with ESO Telescopes at the La Silla Paranal Observatory under programme ID 188.B-3002. These data products have been processed by the Cambridge Astronomy Survey Unit (CASU) at the Institute of Astronomy, University of Cambridge, and by the FLAMES/UVES reduction team at INAF/Osservatorio Astrofisico di Arcetri. These data have been obtained from the Gaia-ESO Survey Data Archive, prepared and hosted by the Wide Field Astronomy Unit, Institute for Astronomy, University of Edinburgh, which is funded by the UK Science and Technology Facilities Council.
This work was partly supported by the European Union FP7 programme through ERC grant number 320360 and by the Leverhulme Trust through grant RPG-2012-541. We acknowledge the support from INAF and Ministero dell' Istruzione, dell' Universit\`a' e della Ricerca (MIUR) in the form of the grant "Premiale VLT 2012". The results presented here benefit from discussions held during the Gaia-ESO workshops and conferences supported by the ESF (European Science Foundation) through the GREAT Research Network Programme. \\

I.S.R. gratefully acknowledges the support provided by the Gemini-CONICYT project 32110029. C.M. acknowledges the support from CONICYT-PCHA/Doctorado Nacional/2014-21141057. D.G. gratefully acknowledges support form the Chilean BASAL Centro de Excelencia en Astrof\'isica y Tecnolog\'ias Afines (CATA) gran PFB-06/2007. S.V. gratefully acknowledges the support provided by Fondecyt reg. 1130721. AK and NK acknowledge the Deutsche Forschungsgemeinschaft for funding from  Emmy-Noether grant  Ko 4161/1. U.H. acknowledges support from the Swedish National Space Board (SNSB). S.G.S acknowledges the support from the Funda\c{c}\~ao para a Ci\^encia e Tecnologia, FCT (Portugal) and POPH/FSE (EC), in the form of the fellowships SFRH/BPD/47611/2008.
\end{acknowledgements}

\nocite{*}
\bibliographystyle{mn2e}
\bibliography{bibliography}

\end{document}